\documentclass[aps,prstab,reprint,groupedaddress,showpacs,amsmath,amssymb]{revtex4-1}
\usepackage{epsfig}
\usepackage{dcolumn}
\usepackage{bm}

\begin{document}

\title{Rotating system for four-dimensional transverse rms-emittance measurements}

l
\author{C.~Xiao, M.~Maier, X.N.~Du, P.~Gerhard, L.~Groening, S.~Mickat, and H.~Vormann}
\affiliation{GSI Helmholtzzentrum f\"ur Schwerionenforschung GmbH, D-64291 Darmstadt, Germany}


\date{\today}

\begin{abstract}
Knowledge of the transverse four-dimensional beam rms-parameters is essential for applications that involve lattice elements that couple the two transverse
degrees of freedom (planes). Of special interest is the elimination of inter-plane correlations to reduce the projected emittances. A dedicated ROtating System for Emittance measurements (ROSE) has been proposed, developed, and successfully commissioned to fully determine the four-dimensional beam matrix. This device has been used at the High Charge injector (HLI) at GSI in a beam line which is composed of a skew quadrupole triplet, a normal quadrupole doublet, and ROSE. Mathematical algorithms, measurements, and results for ion beams of $^{83}$Kr$^{13+}$ at 1.4~MeV/u are reported in this paper.
\end{abstract}

\maketitle


\section{Introduction}
Emittance is an important figure of merit for propagation of charged particle beams. It is defined as the amount of phase-space being occupied by the particle
distribution to quantify the beam quality and to match the following optics. Precise knowledge from measurements of particle distribution parameters is important for accelerator design and for phase-space manipulation. However, most of the published work is just on separated measurements of two-dimensional $x$-$x'$ and $y$-$y'$ sub-phase-spaces (planes)~\cite{Lejeune,Sander,Rimjaem}. For simplicity correlations between the two planes, i.e. $x$-$y$, $x$-$y'$, $x'$-$y$, and $x'$-$y'$ are often assumed as zero. However, such inter-plane correlations may be produced by inter-plane coupling fields such as dipole fringes, solenoids, and titled magnets or just by beam losses~\cite{Lars2}.

Ion beams extracted from Electron Cyclotron Resonance (ECR) ion sources have a complex structure in the four-dimensional phase-space~\cite{Mironov}.
Distributions with equal projected rms-emittances are strongly correlated after extraction~\cite{Peter Spadtke1,Peter Spadtke2,Lars1}. Correlations increase the projected rms-emittances. Removing correlations reduces the effective emittances without introduction of beam loss through scraping for instance.
In order to remove unknown correlations, they must be quantified by measurements. Accordingly, four-dimensional diagnostics has the major task of allowing for
elimination of inter-plane coupling.

For beam lines comprising just non-coupling elements as upright quadrupoles, dipoles, and accelerating gaps, the horizontal and vertical beam dynamics are
decoupled. The design of such beam lines can be accomplished ignoring eventual inter-plane correlations in the beam as long as the projected distribution
parameters are known at the beam line entrance. This convenience of ignoring correlations cannot be further afforded, if the beam line includes coupling
elements as solenoids or skew quadrupoles for instance. In that cases even the projected distribution parameters along the beam line will depend on
inter-plane correlations at the entrance of the beam line~\cite{Chen3}.

Using standard slit/grid emittance measurement devices~\cite{Peter2,Peng} and multi-slit/screen devices~\cite{Catani} as the pepper-pot technique the projected phase-space distributions, i.e. the horizontal and vertical rms-emittances,
can be measured. The slit determines the location of the phase-space element. A subsequent grid measures the angular distribution of the ions that passed the
slit. By moving the slit and recording the angular distribution at each slit position, the projected phases-space distribution is measured. The direction of
movements of the slit and grid determines the plane onto which the four-dimensional distribution is projected. Inter-plane correlation matrix-elements cannot be measured directly using a slit/grid configuration.

There is considerable work on measuring four-dimensional distributions using pepper-pots~\cite{Kondrashev,Kremers,Nagatomo} for ion beams at energies below 150~keV/u, where the beam can be stopped by the pepper-pot mask. However, this method is not applicable at energies beyond 150~keV/u due to technical reasons, i.e. doubtful readout by temperature-dependent screens and fixed resolutions by holes and screens~\cite{Peter}. Four-dimensional emittance measurements were proposed and conducted for instance in~\cite{Thomas,John,Woodley,Prat,Jogren} at electron machines. Other options based on phase space tomography technique have been developed to reconstruct the two-dimensional phase space in~\cite{Lohl} and the full four-dimensional phase space in~\cite{Hock1,Hock2}. The combination of skew quadrupoles with a slit/grid emittance measurement devices has been applied successfully for high intensity uranium ions at an energy of 11.4~MeV/u at GSI~\cite{Chen4}. This paper reports on a method without skew quadrupoles that features reduced time (about a factor of three) needed to perform the measurements within about one hour.

The paper starts with an introduction of the parameters that quantify four-dimensional particle distributions. The third section is on the method
of ROSE: analytical calculation and numerical analysis are elaborated comprehensively. The subsequent section is on showing the commissioning method and software for measuring and evaluating the full four-dimensional beam matrix. The fifth section shows the capability of ROSE to provide the input for successful elimination of inter-plane coupling.
\section{four-dimensional rms-quantities}
Four-dimensional beam rms-emittance measurements require determination of ten unique elements of the second moments beam matrix. The 4$\times$4
symmetric second moments beam matrix $C$ can be expressed as~\cite{TRACE-3D}
\begin{equation}
\label{beam_matrix}
C=
\begin{bmatrix}
\langle xx \rangle &  \langle xx'\rangle &  \langle xy\rangle & \langle xy'\rangle \\
\langle x'x\rangle &  \langle x'x'\rangle & \langle x'y\rangle & \langle x'y'\rangle \\
\langle yx\rangle &  \langle yx'\rangle &  \langle yy\rangle & \langle yy'\rangle \\
\langle y'x\rangle &  \langle y'x'\rangle & \langle y'y\rangle & \langle y'y'\rangle
\end{bmatrix},
\end{equation}
where $x$ and $y$ are the horizontal and vertical coordinates, respectively, and $x'$ and $y'$ are their derivatives with respect to the longitudinal coordinate.

Four of the matrix elements quantify the coupling. If at least one of the elements of the
off-diagonal sub-matrix is non-zero, the beam is transversely coupled. Projected rms-emittances $\varepsilon_x$ and $\varepsilon_y$ are quantities which are used to characterize the transverse beam quality in the laboratory coordinate system and are invariant under linear uncoupled (with respect to the laboratory coordinate system) symplectic transformations. Projected rms-emittances are the rms phase-space areas from projections of the particle distribution onto the planes, and their values are equal to the square roots of the determinants of the on-diagonal 2$\times$2 sub-matrices, i.e. phase-space area divided by~$\pi$:
\begin{equation}
\label{eqdd}
\varepsilon_\mu=\sqrt {\langle \mu\mu \rangle \langle \mu'\mu'\rangle - \langle\mu\mu'\rangle^2},
\end{equation}
where $\mu$ refers to either $x$ or $y$. The dimensionless parameter $\alpha$ relates to the $\mu$-$\mu'$ correlation and the $\beta$-function refers to the beam width. They are defined as
\begin{equation}
\alpha_\mu=-\frac{\langle \mu\mu' \rangle }{\varepsilon_\mu},~~~~~\beta_\mu=\frac{\langle \mu\mu \rangle }{\varepsilon_\mu}.
\end{equation}

The eigen-emittances $\varepsilon_1$ and $\varepsilon_2$ are invariant under coupled linear symplectic transformations provided by solenoids or skew
quadrupoles for instance~\cite{Dragt_pra}. None of the projected emittances can be smaller than the smaller of the two eigen-emittances.
The eigen-emittances can be expressed as~\cite{Chen}
\begin{equation}
\label{eqhh}
{\varepsilon_{1,2}}=\frac{1}{2} \sqrt{-tr[(CJ)^2] \pm \sqrt{tr^2[(CJ)^2]-16 \lvert C \rvert}}.
\end{equation}

The square matrix $J$ is the skew-symmetric matrix with non-zero entries in the block diagonal off form and $J$ is defined as
\begin{equation}
J:=
\begin{bmatrix}
0 &  1 &  0 & 0 \\
-1 &  0 &  0 & 0 \\
0 &  0 &  0 & 1 \\
0 &  0 & -1 & 0
\end{bmatrix}.
\end{equation}

The eigen-emittances are equal to projected rms-emittances if and only if all inter-plane correlations are zero. If the second moments beam matrix has
correlations between horizontal and vertical phase-spaces (see Equ.~\ref{beam_matrix}), the eigen-emittances and projected rms-emittances are different. The product of the eigen-emittances can not be larger than the product of projected rms-emittances. The four-dimensional beam rms-emittance is calculated as
\begin{equation}
\label{eq22}
{\varepsilon_{4d}}=\varepsilon_1\varepsilon_2=\sqrt {\lvert C \rvert} \leqslant \varepsilon_x\varepsilon_y.
\end{equation}

The coupling parameter $t$ is introduced to quantify inter-plane coupling as
\begin{equation}
\label{t value}
t:=\frac{\varepsilon_x \varepsilon_y}{\varepsilon_1 \varepsilon_2}-1 \geqslant 0,
\end{equation}
and if $t$ is equal to zero, there are no inter-plane correlations and the projected rms-emittances are equal to the eigen-emittances.
\section{rose method}
\begin{figure}[hbt]
\centering
\includegraphics*[width=85mm,clip=]{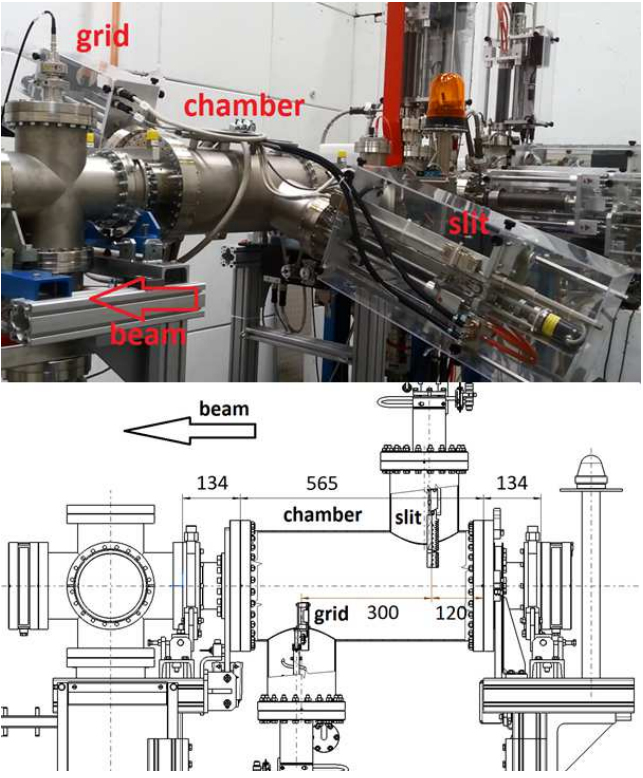}
\caption{ROtating System for Emittance measurements (ROSE): a rotatable chamber with slit and grid actuators being installed at opposite sides to minimize the torque.The total length of ROSE is 833~mm.}
\label{chamber}
\end{figure}

ROSE has been developed to measure the full four-dimensional transverse beam matrix of ion beams as shown in Fig.~\ref{chamber}. It is a slit/grid combination being installed inside a rotatable vacuum chamber. In the slit/grid system of ROSE the slit has an opening width of $d_{slit}$=0.2~mm and the step width is typically $\delta \mu$=0.5~mm. Slit and grid are separated by $d$=300~mm and the wire distance is $d_{wire}$=1.0~mm and the intermediate step number for moving the grid is $n$=4. The spacial/angular resolution of the emittance measurements is accordingly
\begin{equation}
\label{RRR}
\Delta \mu:=\sqrt{(d_{slit})^2+(\delta \mu)^2},
\end{equation}
\begin{equation}
\Delta \mu'=\frac{\sqrt{\left( \frac{d_{wire}}{1+n}\right)^2+(d_{slit})^2}}{d}
\end{equation}
and the spacial/angular resolution of the emittance measurements is determined to be 0.5~mm/0.9~mrad. The emittance measurement unit can be rotated around the beam axis by a total of 270$^{\circ}$, and the rotation can be done within two minutes. One emittance measurement takes about 15 minutes. For rotation no shutters need to be closed. The vacuum pressure increased from few $10^{-8}$~mbar to few $10^{-7}$~mbar during rotation. Afterward the pressure recovered within about three minutes. A detail description of the mechanical set-up can be found in~\cite{ROSE}.

\begin{figure*}[hbt]
\centering
\includegraphics*[width=150mm,clip=]{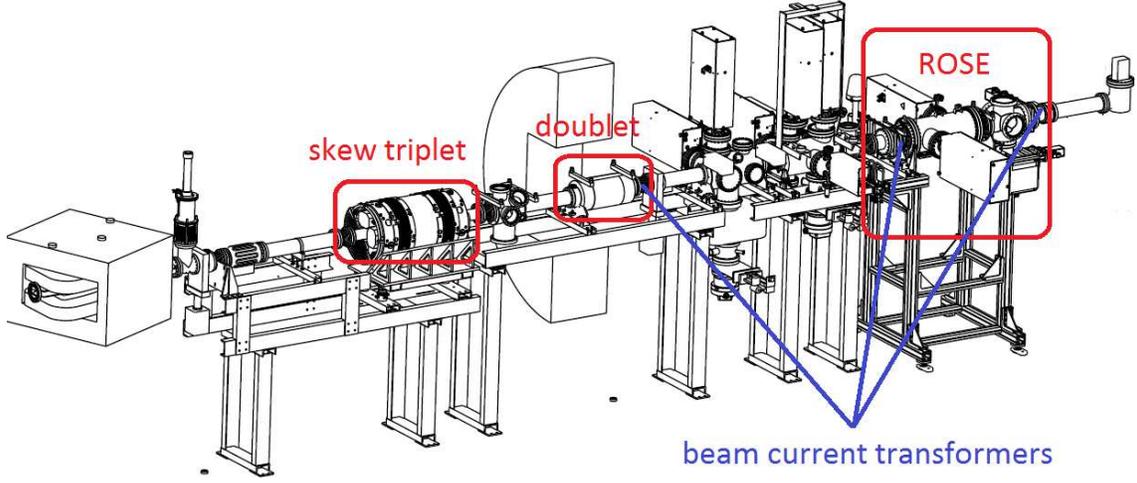}
\caption{Setup of the beam line with ROSE. There are three beam current transformers in this section, the first one is behind a quadrupole doublet, the second one is before ROSE, and the third one is behind ROSE. The beam enters from the left.}
\label{ROSE_beamline}
\end{figure*}

ROSE has been installed at the HLI section~\cite{Barth}, as it is fed by an ECR source that provides correlated beams. It is installed as a mobile set-up, i.e. the corresponding chamber may be installed at many locations along the versatile GSI beam lines. The complete beam line consists of one skew quadrupole triplet and one regular quadrupole doublet followed by the ROSE unit as shown in Fig.~\ref{ROSE_beamline}.
\subsection{Mathematical algorithms}
\begin{figure}[hbt]
\centering
\includegraphics*[width=85mm,clip=]{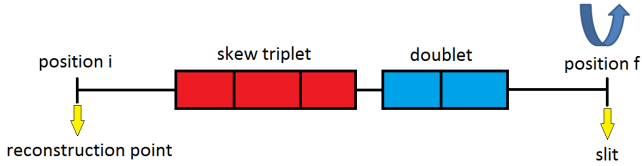}
\caption{Sketch of the ROSE beam line with a rotatable slit/grid emittance scanner. All beam second moments are measured at location $f$. Applying the inverted transfer matrix they are transported backward to location $i$ of the initial and constant beam matrix $C_i$. Full beam transmission between location $i$ and location $f$ is required.}
\label{ROSE_beamline1}
\end{figure}

The transport of the beam matrix from location $i$ to location $f$ can be calculated as (see Fig.~\ref{ROSE_beamline1})
\begin{equation}
\label{transport}
C_f=M C_i M^T,
\end{equation}
where $M$ is the transport matrix between location $i$ and location $f$
\begin{equation}
\label{transport11}
M=
\begin{bmatrix}
m_{11} & m_{12} &  m_{13} &  m_{14} \\
m_{21} & m_{22} &  m_{23} &  m_{24} \\
m_{31} & m_{32} &  m_{33} & m_{34} \\
m_{41} & m_{42} &  m_{43} & m_{44}
\end{bmatrix}
:=
\begin{bmatrix}
M_{xx} & M_{xy}\\
M_{yx} & M_{yy}
\end{bmatrix}.
\end{equation}

As the ROSE beam line is without coupling elements if the skew quadrupole triplet is switched off, the off-diagonal sub-matrices of the transport matrices
vanish, i.e. $M_{xy}$=$M_{yx}$=0. The transports $M^a$ or $M^b$ of single particle coordinates from location $i$ to location $f$ using magnet
setting $a$ or $b$ are described by transfer matrices through
\begin{equation}
\label{transport22}
\begin{bmatrix}
x\\
x'\\
y\\
y'
\end{bmatrix}^{a,b}_f
=
\begin{bmatrix}
m^{a,b}_{11} & m^{a,b}_{12} & 0 & 0 \\
m^{a,b}_{21} & m^{a,b}_{22} & 0 & 0 \\
0 & 0 & m^{a,b}_{33} & m^{a,b}_{34} \\
0 & 0 & m^{a,b}_{43} & m^{a,b}_{44} \,.
\end{bmatrix}
\begin{bmatrix}
x\\
x'\\
y\\
y'
\end{bmatrix}_i,
\end{equation}
from Equ.~\ref{transport22}, the correlated beam second moments of the off-diagonal sub-matrices at location $f$ using magnet settings $a$ and $b$ can be
written as
\begin{equation}
\label{coupling1}
\begin{aligned}
&\langle xy \rangle^{a,b}_f=m^{a,b}_{11}m^{a,b}_{33}\langle xy \rangle_i+m^{a,b}_{11}m^{a,b}_{34}\langle xy' \rangle_i
\\
&+m^{a,b}_{12}m^{a,b}_{33}\langle x'y \rangle_i+m^{a,b}_{12}m^{a,b}_{34}\langle x'y' \rangle_i,
\end{aligned}
\end{equation}
\begin{equation}
\label{coupling2}
\begin{aligned}
&\langle xy' \rangle^{a,b}_f+\langle x'y \rangle^{a,b}_f=(m^{a,b}_{11}m^{a,b}_{43}+m^{a,b}_{21}m^{a,b}_{33})\langle xy \rangle_i\\
&+(m^{a,b}_{11}m^{a,b}_{44}+m^{a,b}_{21}m^{a,b}_{34})\langle xy' \rangle_i\\
&+(m^{a,b}_{12}m^{a,b}_{43}+m^{a,b}_{22}m^{a,b}_{33})\langle x'y \rangle_i\\
&+(m^{a,b}_{12}m^{a,b}_{44}+m^{a,b}_{22}m^{a,b}_{34})\langle x'y' \rangle_i,
\end{aligned}
\end{equation}
\begin{equation}
\label{coupling3}
\begin{aligned}
&\langle x'y' \rangle^{a,b}_f=m^{a,b}_{21}m^{a,b}_{43}\langle xy \rangle_i+m^{a,b}_{21}m^{a,b}_{44}\langle xy' \rangle_i
\\
&+m^{a,b}_{22}m^{a,b}_{43}\langle x'y \rangle_i+m^{a,b}_{22}m^{a,b}_{44}\langle x'y' \rangle_i.
\end{aligned}
\end{equation}

Rotating clockwise the emittance measurement unit by $\theta$ is equivalent to rotating the beam by -$\theta$ around the beam axis. After rotation, the
new particle coordinates using magnet settings $a$ or $b$ are transported by a simple rotation matrix
\begin{equation}
\label{transport3}
\begin{bmatrix}
x\\
x'\\
y\\
y'
\end{bmatrix}^{a,b}_{\theta}
=
\begin{bmatrix}
cos\theta &0 & sin\theta & 0 \\
0 & cos\theta & 0 & sin\theta \\
-sin\theta & 0 & cos\theta & 0 \\
0 & -sin\theta & 0 & cos\theta
\end{bmatrix}
\begin{bmatrix}
x\\
x'\\
y\\
y'
\end{bmatrix}^{a,b}_f.
\end{equation}

According to Equ.~\ref{transport3} horizontal second moments after rotation using magnet settings $a$ and $b$ are expressed as
\begin{equation}
\label{xx'}
\begin{aligned}
&\langle xx \rangle^{a,b}_\theta=cos^2\theta\langle xx \rangle^{a,b}_f+2sin\theta cos\theta \langle xy \rangle^{a,b}_f\\
&+sin^2\theta \langle yy \rangle^{a,b}_f,
\end{aligned}
\end{equation}
\begin{equation}
\label{xx'}
\begin{aligned}
&\langle xx' \rangle^{a,b}_\theta=cos^2\theta\langle xx' \rangle^{a,b}_f+sin\theta cos\theta \langle xy' \rangle^{a,b}_f\\
&+sin\theta cos\theta \langle x'y \rangle^{a,b}_f+sin^2\theta \langle yy' \rangle^{a,b}_f,
\end{aligned}
\end{equation}
\begin{equation}
\label{xx' a and b}
\begin{aligned}
&\langle x'x' \rangle^{a,b}_\theta=cos^2\theta\langle x'x' \rangle^{a,b}_f+2sin\theta cos\theta \langle x'y' \rangle^{a,b}_f\\
&+sin^2\theta \langle y'y' \rangle^{a,b}_f.
\end{aligned}
\end{equation}

All elements of the transport matrices $M^{a,b}_{xx}$ and $M^{a,b}_{yy}$ are known. The second moments $\langle xx \rangle^{a,b}_f$, $\langle xx' \rangle^{a,b}_f$,
$\langle x'x' \rangle^{a,b}_f$, $\langle yy \rangle^{a,b}_f$, $\langle yy' \rangle^{a,b}_f$, and $\langle y'y' \rangle^{a,b}_f$ before rotation and
$\langle xx \rangle^{a,b}_\theta$, $\langle xx' \rangle^{a,b}_\theta$, and $\langle x'x' \rangle^{a,b}_\theta$ after rotation can be measured.
Combining Equ.~\ref{coupling1} to Equ.~\ref{xx' a and b} the solution of the searched coupled matrix elements at location $i$ can be summarized to a set of
linear equations
\begin{equation}
\label{coupling001}
\begin{cases}
\begin{aligned}
&\Gamma_{11}\langle xy \rangle_i+\Gamma_{12}\langle xy' \rangle_i+\Gamma_{13}\langle x'y \rangle_i+\Gamma_{14}\langle x'y' \rangle_i=\Lambda_1\\
&\Gamma_{21}\langle xy \rangle_i+\Gamma_{22}\langle xy' \rangle_i+\Gamma_{23}\langle x'y \rangle_i+\Gamma_{24}\langle x'y' \rangle_i=\Lambda_2\\
&\Gamma_{31}\langle xy \rangle_i+\Gamma_{32}\langle xy' \rangle_i+\Gamma_{33}\langle x'y \rangle_i+\Gamma_{34}\langle x'y' \rangle_i=\Lambda_3\\
&\Gamma_{41}\langle xy \rangle_i+\Gamma_{42}\langle xy' \rangle_i+\Gamma_{43}\langle x'y \rangle_i+\Gamma_{44}\langle x'y' \rangle_i=\Lambda_4\\
&\Gamma_{51}\langle xy \rangle_i+\Gamma_{52}\langle xy' \rangle_i+\Gamma_{53}\langle x'y \rangle_i+\Gamma_{54}\langle x'y' \rangle_i=\Lambda_5\\
&\Gamma_{61}\langle xy \rangle_i+\Gamma_{62}\langle xy' \rangle_i+\Gamma_{63}\langle x'y \rangle_i+\Gamma_{64}\langle x'y' \rangle_i=\Lambda_6
\end{aligned}
\end{cases}
\end{equation}
with
\begin{equation}
\label{xx'}
\begin{aligned}
&\Gamma_{11}=m^a_{11}m^a_{33},~\Gamma_{12}=m^a_{11}m^a_{34},~\Gamma_{13}=m^a_{12}m^a_{33},\\
&\Gamma_{14}=m^a_{12}m^a_{34},
\end{aligned}
\end{equation}
\begin{equation}
\begin{aligned}
&\Gamma_{21}=m^a_{11}m^a_{43}+m^a_{21}m^a_{33},~\Gamma_{22}=m^a_{11}m^a_{44}+m^a_{21}m^a_{34},\\
&\Gamma_{23}=m^a_{12}m^a_{43}+m^a_{22}m^a_{33},~\Gamma_{24}=m^a_{12}m^a_{44}+m^a_{22}m^a_{34},\\
\end{aligned}
\end{equation}
\begin{equation}
\begin{aligned}
&\Gamma_{31}=m^a_{21}m^a_{43},~\Gamma_{32}=m^a_{21}m^a_{44},~\Gamma_{33}=m^a_{22}m^a_{43},\\
&\Gamma_{34}=m^a_{22}m^a_{44},\\
\end{aligned}
\end{equation}
and with the same procedure for setting $b$
\begin{equation}
\begin{aligned}
&\Gamma_{41}=m^b_{11}m^b_{33},~\Gamma_{42}=m^b_{11}m^b_{34},~\Gamma_{43}=m^b_{12}m^b_{33},\\
&\Gamma_{44}=m^b_{12}m^b_{34},
\end{aligned}
\end{equation}
\begin{equation}
\begin{aligned}
&\Gamma_{51}=m^b_{11}m^b_{43}+m^b_{21}m^b_{33},~\Gamma_{52}=m^b_{11}m^b_{44}+m^b_{21}m^b_{34},\\
&\Gamma_{53}=m^b_{12}m^b_{43}+m^b_{22}m^b_{33},~\Gamma_{54}=m^b_{12}m^b_{44}+m^b_{22}m^b_{34},\\
\end{aligned}
\end{equation}
\begin{equation}
\begin{aligned}
&\Gamma_{61}=m^b_{21}m^b_{43},~\Gamma_{62}=m^b_{21}m^b_{44},~\Gamma_{63}=m^b_{22}m^b_{43},\\
&\Gamma_{64}=m^b_{22}m^b_{44},\\
\end{aligned}
\end{equation}
and finally
\begin{equation}
\label{xx'}
\begin{aligned}
\Lambda_{1}=\langle xy \rangle^a_f, ~\Lambda_{2}=\langle xy' \rangle^a_f+\langle x'y \rangle^a_f, ~\Lambda_{3}=\langle x'y' \rangle^a_f,\\
\end{aligned}
\end{equation}
\begin{equation}
\begin{aligned}
\Lambda_{4}=\langle xy \rangle^b_f, ~\Lambda_{5}=\langle xy' \rangle^b_f+\langle x'y \rangle^b_f, ~\Lambda_{6}=\langle x'y' \rangle^b_f.\\
\end{aligned}
\end{equation}

Since there are $k$=4 unknown coupling parameters and $n$=6 linear equations, at least $k$=4 of them are selected to solve the unknown parameters. There are
\begin{equation}
\label{zhuhe}
\begin{aligned}
C^k_n
=\frac{n!}{k!(n-k)!}=15
\end{aligned}
\end{equation}
possibilities to do so. From Equ.~\ref{coupling001} two arbitrary equations can be removed and doing so in total 15 algorithms are generated.
For instance algorithm~$\#$1 reads
\begin{equation}
\label{coupling01}
\begin{bmatrix}
\langle xy \rangle_i \\
\langle xy' \rangle_i\\
\langle x'y \rangle_i \\
\langle x'y' \rangle_i
\end{bmatrix}^1=
\begin{bmatrix}
\Gamma_{11} & \Gamma_{12} & \Gamma_{13} & \Gamma_{14} \\
\Gamma_{21} & \Gamma_{22} & \Gamma_{23} & \Gamma_{24} \\
\Gamma_{31} & \Gamma_{32} & \Gamma_{33} & \Gamma_{34} \\
\Gamma_{41} & \Gamma_{42} & \Gamma_{43} & \Gamma_{44}
\end{bmatrix}^{-1}
\begin{bmatrix}
\Lambda_1 \\
\Lambda_2 \\
\Lambda_3 \\
\Lambda_4
\end{bmatrix},
\end{equation}
and algorithm~$\#$15 reads
\begin{equation}
\label{coupling15}
\begin{bmatrix}
\langle xy \rangle_i \\
\langle xy' \rangle_i\\
\langle x'y \rangle_i \\
\langle x'y' \rangle_i
\end{bmatrix}^{15}=
\begin{bmatrix}
\Gamma_{31} & \Gamma_{32} & \Gamma_{33} & \Gamma_{34} \\
\Gamma_{41} & \Gamma_{42} & \Gamma_{43} & \Gamma_{44} \\
\Gamma_{51} & \Gamma_{52} & \Gamma_{53} & \Gamma_{54} \\
\Gamma_{61} & \Gamma_{62} & \Gamma_{63} & \Gamma_{64}
\end{bmatrix}^{-1}
\begin{bmatrix}
\Lambda_3 \\
\Lambda_4 \\
\Lambda_5 \\
\Lambda_6
\end{bmatrix}.
\end{equation}

The algorithms offer the possibility to reconstruct the full beam matrix at location $i$ from emittance measurements of different rotation angles through
\begin{equation}
\label{final}
\begin{bmatrix}
\langle xy \rangle_i \\
\langle xy' \rangle_i\\
\langle x'y \rangle_i \\
\langle x'y' \rangle_i
\end{bmatrix}^j
=
\Gamma^{-1}_j \Lambda^j,~~~j=1,2,\cdots 15.
\end{equation}
\subsection{Measurement procedure}
Projected rms-emittance measurements can be performed at various angles, i.e. 0$^{\circ}$, 90$^{\circ}$, and $\Theta$$^{\circ}$ (any angle which is not
equivalent to 0$^{\circ}$ or 90$^{\circ}$) to reconstruct the full four-dimensional beam matrix $C_i$. Rotation by 0$^{\circ}$/90$^{\circ}$ will just measure
the usual uncoupled second moments. Rotation by $\Theta$$^{\circ}$ provides access to the coupled beam second moments. As will be shown below, this optics
comprises just non-coupling linear elements. Using a rotatable emittance measurement device, a minimum of four, but more reliable six measurements is sufficient
to measure the complete four-dimensional beam matrix:
\\

(\uppercase\expandafter{\romannumeral1}) measurements at $\theta$=0$^{\circ}$ with optics $a$ (and $b$).

(\uppercase\expandafter{\romannumeral2}) measurements at $\theta$=90$^{\circ}$ with optics $a$ (and $b$).

(\uppercase\expandafter{\romannumeral3}) measurements at $\theta$=$\Theta$$^{\circ}$ with optics $a$ and $b$.
\\

If just four measurements are applied to evaluate the full beam matrix at location $i$, the uncoupled second moments for setting $b$ at
location $f$, i.e. $\langle xx \rangle^b_f$, $\langle xx' \rangle^b_f$, and $\langle x'x' \rangle^b_f$ are calculated from the final uncoupled second moments
for setting $a$, i.e. $\langle xx \rangle^a_f$, $\langle xx' \rangle^a_f$, and $\langle x'x' \rangle^a_f$ using the transport matrices $M^a_{xx}$ and $M^b_{xx}$.

The four measurements (projected rms-emittances and Twiss parameters) and their deliverables are:
\\

(\uppercase\expandafter{\romannumeral1}) $\theta$=0$^{\circ}$, magnet setting $a$ delivers parameters $\langle xx \rangle^a_f$, $\langle xx' \rangle^a_f$,
and $\langle x'x' \rangle^a_f$.

(\uppercase\expandafter{\romannumeral2}) $\theta$=90$^{\circ}$, magnet setting $a$ delivers parameters $\langle yy \rangle^a_f$, $\langle yy' \rangle^a_f$,
and $\langle y'y' \rangle^a_f$.

(\uppercase\expandafter{\romannumeral3}) $\theta$=$\Theta$$^{\circ}$, magnet setting $a$ delivers parameters $\langle xx \rangle^a_{\theta}$,
$\langle xx' \rangle^a_{\theta}$, and $\langle x'x' \rangle^a_{\theta}$.

(\uppercase\expandafter{\romannumeral4}) $\theta$=$\Theta$$^{\circ}$, magnet setting $b$ delivers parameter $\langle xx \rangle^b_{\theta}$,
$\langle xx' \rangle^b_{\theta}$, and $\langle x'x' \rangle^b_{\theta}$.
\\

From step (\uppercase\expandafter{\romannumeral1}) the uncorrelated second moments $\langle xx \rangle_i$, $\langle xx' \rangle_i$, and
$\langle x'x' \rangle_i$ are obtained at location $i$ by simple back transformation through inversion of Equ.~\ref{transport22}.
From step (\uppercase\expandafter{\romannumeral2}) the uncorrelated beam-moments $\langle yy \rangle_i$, $\langle yy' \rangle_i$, and
$\langle y'y' \rangle_i$ are obtained at location $i$ in the same way. Steps (\uppercase\expandafter{\romannumeral3})
and (\uppercase\expandafter{\romannumeral4}) deliver $\langle xy \rangle^{a,b}_f$, $\langle xy' \rangle^{a,b}_f$+$\langle x'y \rangle^{a,b}_f$,
and $\langle x'y' \rangle^{a,b}_f$ at location $f$ (Equ.~\ref{coupling1} to Equ.~\ref{coupling3}). Finally, Equ.~\ref{final} determines the
correlated second moments $\langle xy \rangle_i$, $\langle xy' \rangle_i$, $\langle x'y \rangle_i$, and $\langle x'y' \rangle_i$ at location $i$.
The four-dimensional second moments beam matrix is then finally reconstructed at location $i$ from four measurements.

If six measurements are applied, the uncoupled second moments for setting $b$ at location $f$ can be measured directly.
\subsection{Minimizing the measurement errors}
The vector $\Lambda^j$ (see Equ.~\ref{final}) is sensitive to the emittance measurements at location $f$. During emittance measurements,
finite grid bin results in finite resolution and background noise have influence on the measured second moments. The typical error of directly measured second moments is about 10$\%$. These errors enter into the inversion
of Equ.~\ref{final}.

In the following the minimization of the measurement error of the coupled second moments by making use of the so-called condition number of a matrix is described. If the inverse ${\Gamma_j}^{-1}$
exists, the condition number of a square matrix $\Gamma_j$ is defined as
\begin{equation}
\label{cond}
\begin{aligned}
\kappa_j(\Gamma_j):={\Vert \Gamma_j \Vert}_2 {\Vert \Gamma_j^{-1} \Vert}_2.
 \end{aligned}
\end{equation}

This quantity is always bigger than or equal to 1.0. Let $\Gamma_j^T$ be the transpose of the square matrix $\Gamma_j$, then the spectral norm, or two-norm,
of a matrix is defined as the square root of the maximum eigenvalue of $\Gamma_j^T \Gamma_j$
\begin{equation}
\begin{aligned}
{\Vert \Gamma_j \Vert}_2:= \sqrt{\rho(\Gamma_j^T \Gamma_j)},
 \end{aligned}
\end{equation}
where $\rho$ is the function that computes the spectral radius of $\Gamma_j$.
Since the matrix $\Gamma_j^T \Gamma_j$ is symmetric and all of its eigenvalues are real-valued and non-negative then $\rho(\Gamma_j^T \Gamma_j)$ is the
largest of these eigenvalues.

The square matrix $\Gamma_j$, related to the transfer matrix-elements from location $i$ to location $f$, is invertible but ill-conditioned if its condition number
is too large. The condition number associated with the linear equations (see Equ.~\ref{final}) is a measure for how ill-conditioned the matrix is. If the
condition number is large, even a small error in emittance measurements may lead to radically different results for the beam coupling parameter evaluations.
On the other hand, if the condition number is small the error in evaluation will not exceed notably the error in emittance measurements.
The numerical stability (degeneration of the system) is better if the condition number is small. Well-conditioned matrices have condition numbers which are
closed to 1.0.

We summarize that in order to obtain reasonable evaluation results it needs:

(\uppercase\expandafter{\romannumeral1}) one reference emittance measurement with 100$\%$ transmission efficiency between location $i$ and location $f$ to
obtain projected beam parameters at location $i$ (on-diagonal section of beam matrix of $C_i$).

(\uppercase\expandafter{\romannumeral2}) that all quadrupoles are varied numerically in a brute-force method to check each setting for full transmission efficiency
from location $i$ to location $f$, and for reasonable beam sizes on slit/grid ( $2$~mm$<\sigma_{rms}<$$5$~mm in our case). In the plane spanned by the two
quadrupole gradients these settings form finite areas. We refer to these areas as safety islands in the following.

(\uppercase\expandafter{\romannumeral3}) that all settings from safety islands are combined to determine combinations of two settings $a$ and $b$ corresponding to a
low condition number.
\section{measurements and evaluations}
In a first measurement the projected rms-emittances and Twiss parameters at the exit of the ROSE beam line were measured as listed in Tab.~\ref{tab_3}.
A beam of $^{83}$Kr$^{13+}$ at 1.4~MeV/u has been used, the beam intensity through the ROSE was about 20~e$\mu$A, and space-charge effects can be neglected in this case. The skew triplet and normal doublet were switched off and the transmission through the set-up
was 100$\%$.
\begin{table}
\caption{\label{tab_3} Projected rms-emittances and Twiss parameters measured at ROSE for $^{83}$Kr$^{13+}$ at 1.4~MeV/u (reference measurements).}
\begin{ruledtabular}
\begin{tabular}{c|c|c|c}
Rotation angle & $\alpha_{rms}$ & $\beta_{rms}$ [m/rad]& $\varepsilon_{rms}$ [mm~mrad]\\
\hline
0$^{\circ}$ & -2.9 & 12.9 & 2.0\\
90$^{\circ}$ & -1.6 & 10.5 & 2.4	
\end{tabular}
\end{ruledtabular}
\end{table}
Uncoupled second moments at location $i$ were obtained from 0$^{\circ}$/90$^{\circ}$ measurements at location $f$ to reconstruct the on-diagonal section of $C_i$.
In order to match reasonable beam sizes on the slit/gird locations and to assure full transmission, the strengths of $Q_1$ and $Q_2$ were varied numerically to
check all available doublet settings, i.e. the safety islands including all reasonable doublet settings were defined. Combining two settings of the doublet
($Q_1$ and $Q_2$, $Q_1'$ and $Q_2'$) from the safety islands, the corresponding condition number of the matrix $\Gamma_j$ was calculated using Equ.~\ref{cond}.
Since there are $N$ doublet settings inside the safety islands, $N^2$ combinations of two doublet settings were obtained. Finally, the combination of doublet
settings $a$ and $b$ with minimum condition number of the matrix $\Gamma_j$ is applied.

\begin{figure}[hbt]
\centering
\includegraphics*[width=85mm,clip=]{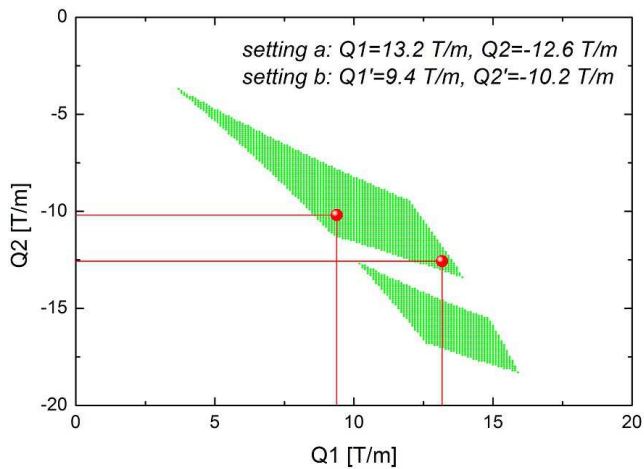}
\caption{Safety islands (green dots) and the selected doublet settings (red dots) applied during the measurements.}
\label{safes}
\end{figure}

The doublet setting~$a$ $Q1/Q2$=13.2/-12.6~T/m and setting~$b$ $Q1'/Q2'$=9.4/-10.2~T/m were selected as they provide low condition numbers for the majority
of the algorithms. The safety islands and the selected doublet settings are plotted in Fig.~\ref{safes}.
\subsection{Beam with low coupling}
\begin{figure}[hbt]
\centering
\includegraphics*[width=85mm,clip=]{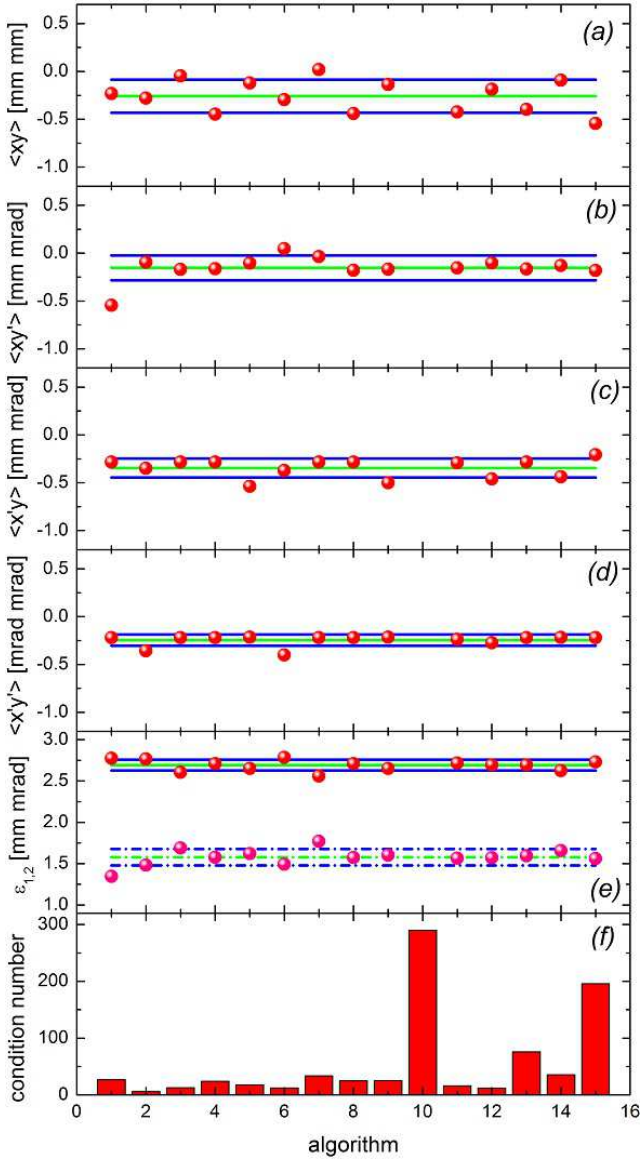}
\caption{Reconstructed results at location~$i$ applying four measurements with the skew triplet being switched off. The green line indicates the mean value and the blue lines indicate the $\pm\sigma$ error range.
(a) to (d): Coupled second moments: $\langle xy \rangle$=-0.3$\pm$0.2~mm~mm, $\langle xy' \rangle$=-0.2$\pm$0.1~mm~mrad, $\langle x'y \rangle$=-0.4$\pm$0.1~mm~mrad, and $\langle x'y' \rangle$=-0.3$\pm$0.1~mrad~mrad.
(e): Eigen-emittances: $\varepsilon_1$=2.7$\pm$0.1~mm~mrad, and $\varepsilon_2$=1.6$\pm$0.1~mm~mrad. (f): Condition numbers for each algorithm. The results of algorithm~$\#{10}$ are considered as unreliable for the large condition number.}
\label{low_coupling_4}
\end{figure}

Measurements at 0$^{\circ}$, 90$^{\circ}$, and -30$^{\circ}$ using settings $a$ and $b$ with the skew triplet being switched off were done.
The measured Twiss parameters together with the projected rms-emittances are listed in Tab.~\ref{tab_4}. Two evaluations using four or six measurements
were done independently for comparison.
\begin{table}[hbt]
\caption{\label{tab_4}Measured projected rms-emittances and Twiss parameters at the exit of the ROSE beam line with the skew triplet being switched off.}
\begin{ruledtabular}
\begin{tabular}{c|c|c|c|c}
Rotation angle & setting &$\alpha_{rms}$ & $\beta_{rms}$ [m/rad]& $\varepsilon_{rms}$ [mm~mrad]\\
\hline
0$^{\circ}$ & a &0.3 & 4.0 & 1.9\\
0$^{\circ}$ & b &-0.0 & 4.5 & 1.9\\
90$^{\circ}$& a  &-1.8 &4.0 & 2.5\\
90$^{\circ}$& b  &-1.1 &6.6 & 2.8\\
-30$^{\circ}$& a &-0.4 &3.8 & 2.3\\
-30$^{\circ}$& b &-0.6 &4.8 & 2.2
\end{tabular}
\end{ruledtabular}
\end{table}
\\
Evaluating four measurements (setting $a$ at 0$^{\circ}$/90$^{\circ}$/-30$^{\circ}$ and setting $b$ at -30$^{\circ}$)
the correlated second moments at location $i$, their corresponding eigen-emittances, and condition number for each algorithm are
shown in Fig.~\ref{low_coupling_4}. The averaged beam second moments matrix $\widehat{C}(4)$ applying four measurements at location~$i$ is calculated as (in units of mm and mrad)
\begin{equation}
\label{overlineC_i}
\widehat{C}(4)=
\begin{bmatrix}
4.8 &  -2.5 & -0.3 & -0.2 \\
-2.5 &  2.1 &  -0.3 & -0.2 \\
 -0.3 & -0.3  & 5.0 & 0.0 \\
-0.2 &   -0.2 & 0.0 & 1.2
\end{bmatrix}.
\end{equation}

\begin{figure}[hbt]
\centering
\includegraphics*[width=85mm,clip=]{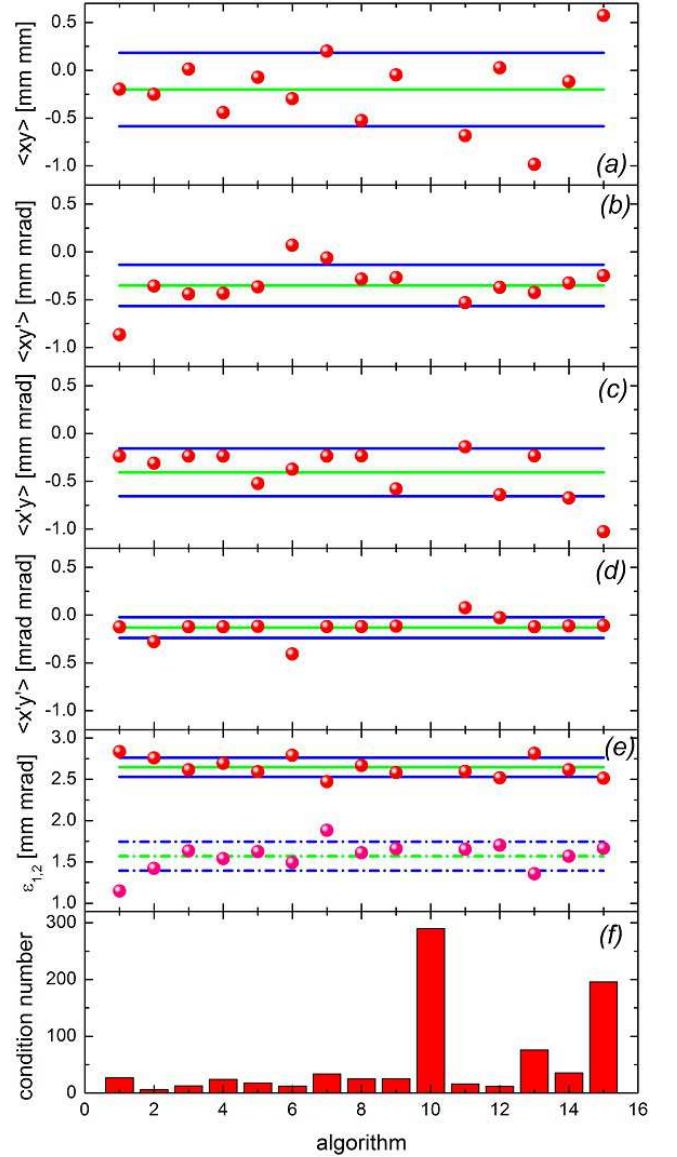}
\caption{Reconstructed results at location~$i$ applying six measurements with the skew triplet being switched off. The green line indicates the mean value and the blue lines indicate the $\pm\sigma$ error range.
(a) to (d): Coupled second moments: $\langle xy \rangle$=-0.2$\pm$0.4~mm~mm, $\langle xy' \rangle$=-0.4$\pm$0.2~mm~mrad, $\langle x'y \rangle$=-0.4$\pm$0.3~mm~mrad, and $\langle x'y' \rangle$=-0.1$\pm$ 0.1~mrad~mrad.
(e): Eigen-emittances: $\varepsilon_1$=2.7$\pm$0.1~mm~mrad, and $\varepsilon_2$=1.6$\pm$0.2~mm~mrad. (f): Condition numbers for each algorithm. The results of algorithm~$\#{10}$ are considered as unreliable for the large condition number.}
\label{low_coupling_6}
\end{figure}

Evaluating six measurements (setting $a$ and $b$ at 0$^{\circ}$/90$^{\circ}$/-30$^{\circ}$) the correlated second moments at location~$i$, their corresponding
eigen-emittances, and condition number for each algorithm are shown in Fig.~\ref{low_coupling_6}. The averaged beam second moments matrix $\widehat{C}(6)$
applying six measurements at location~$i$ is calculated as (in units of mm and mrad)
\begin{equation}
\label{overlineC_i}
\widehat{C}(6)=
\begin{bmatrix}
4.8 &  -2.5 & -0.2 & -0.4 \\
-2.5 &  2.1 &  -0.4 & -0.1 \\
-0.2 & -0.4  & 5.0 & 0.0 \\
-0.4 &   -0.1 & 0.0 & 1.2
\end{bmatrix}.
\end{equation}

\begin{figure}[hbt]
\centering
\includegraphics*[width=85mm,clip=]{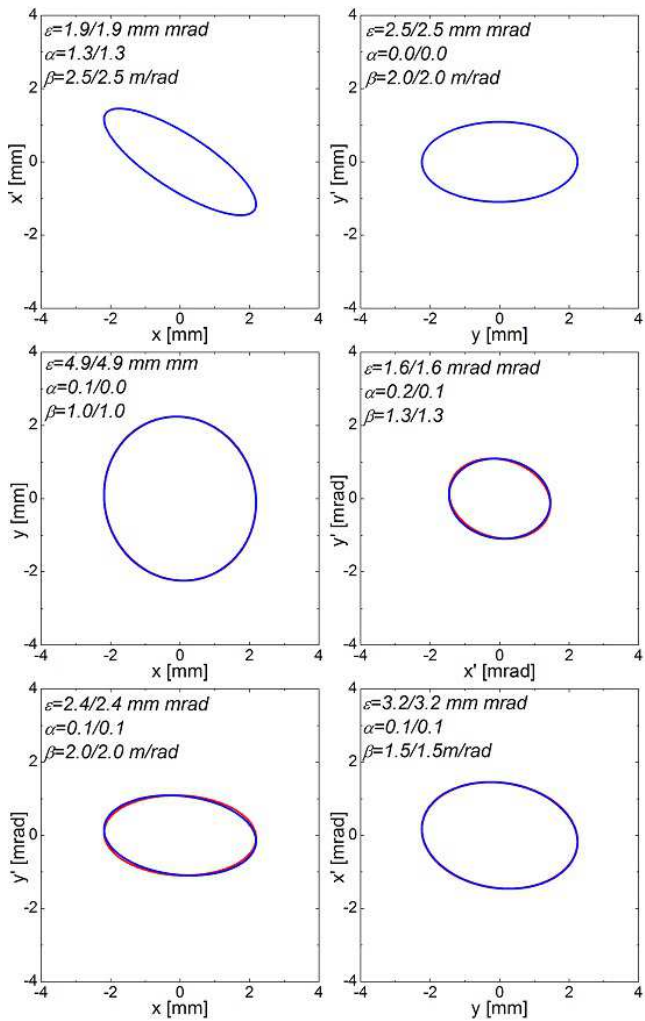}
\caption{Projected rms-ellipses from measurements applying four/six measurements with the skew triplet being switched off. The red and blue ellipses indicate the beam matrices $\widehat{C}(4)$ (from four measurements) and
$\widehat{C}(6)$ (from six measurements). The projected rms-emittances and the Twiss parameters are indicated. The two matrices produce almost identical rms-ellipses.}
\label{coupling_small_c}
\end{figure}

Evaluation of the two eigen-emittances of $\widehat{C}(4)$/$\widehat{C}(6)$ reveals $\varepsilon_1$=2.7/2.6~mm~mrad and
$\varepsilon_2$=1.6/1.6~mm~mrad. The corresponding coupling parameters $t$ are 0.1/0.1. Both evaluations produce similar eigen-emittances and coupling
parameters. The rms-ellipses of the matrices $\widehat{C}(4)$/$\widehat{C}(6)$ in the projections are shown in Fig.~\ref{coupling_small_c}.

\subsection{Beam with large coupling}
In order to create the correlation the skew triplet was switched on. Measurements were done at 0$^{\circ}$, 90$^{\circ}$, and 30$^{\circ}$
using settings $a$ and $b$. The Twiss parameters together with the projected rms-emittances are listed in Tab.~\ref{tab_5}.
\begin{table}
\caption{\label{tab_5}Measured projected rms-emittances and Twiss parameters at the exit of the ROSE beam line with the skew triplet being switched on.}
\begin{ruledtabular}
\begin{tabular}{c|c|c|c|c}
Rotation angle & setting &$\alpha_{rms}$ & $\beta_{rms}$ [m/rad]& $\varepsilon_{rms}$ [mm~mrad]\\
\hline
0$^{\circ}$ & b &-0.1& 4.6 & 3.2\\
0$^{\circ}$ & a &0.0 & 4.0 & 3.1\\
90$^{\circ}$& b &-2.5 &8.8 & 3.4\\
90$^{\circ}$& a &-2.7 &7.4 & 3.3\\
30$^{\circ}$& b &-0.6 &2.3 &3.2\\
30$^{\circ}$& a &-0.8 &2.7 &4.7
\end{tabular}
\end{ruledtabular}
\end{table}

\begin{figure}[hbt]
\centering
\includegraphics*[width=85mm,clip=]{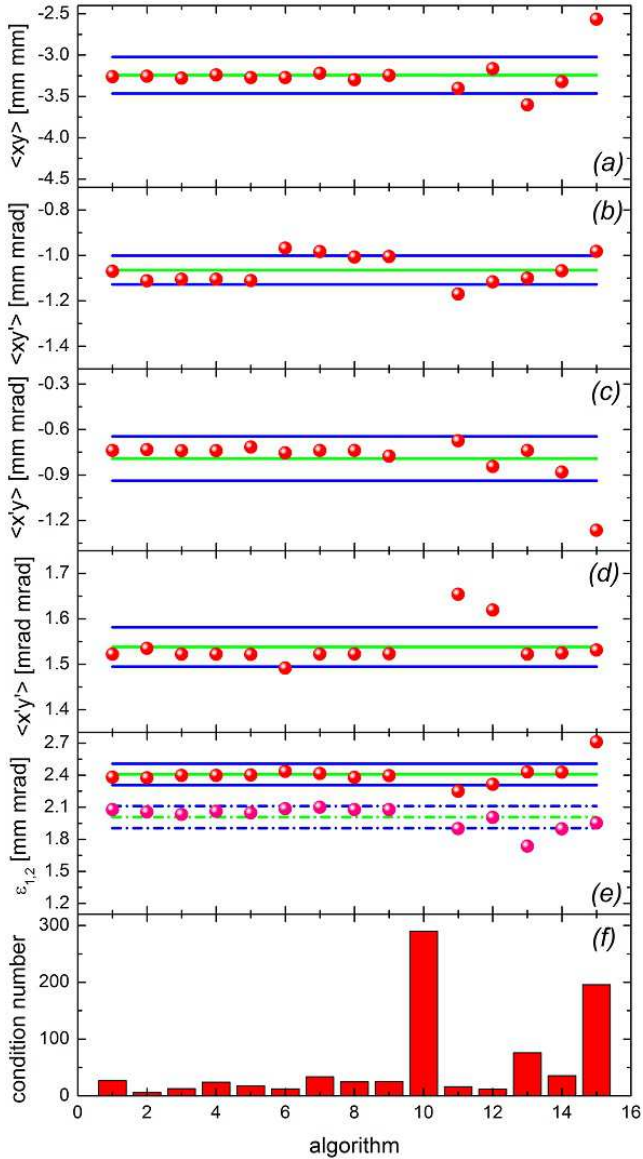}
\caption{Reconstructed results at location~$i$ applying four measurements with the skew triplet being switched on. The green line indicates the mean value and the blue lines indicate the $\pm\sigma$ error range.
(a) to (d): Coupled second moments: $\langle xy \rangle$=-3.2$\pm$0.2~mm~mm, $\langle xy' \rangle$=-1.1$\pm$0.1~mm~mrad, $\langle x'y \rangle$=-0.8$\pm$0.2~mm~mrad, and $\langle x'y' \rangle$=1.5$\pm$0.0~mrad~mrad.
(e): Eigen-emittances: $\varepsilon_1$=2.4$\pm$0.1~mm~mrad, and $\varepsilon_2$=2.0$\pm$0.1~mm~mrad. (f): Condition numbers for each algorithm. The results of algorithm~$\#{10}$ are considered as unreliable for the large condition number.}
\label{large_coupling_4}
\end{figure}

The 15 evaluations of correlated second moments at location~$i$, their corresponding
eigen-emittances, and their condition number for each algorithm using four measurements (setting $b$ at 0$^{\circ}$/90$^{\circ}$/30$^{\circ}$ and
setting $a$ at 30$^{\circ}$) are shown in Fig.~\ref{large_coupling_4}.
The corresponding beam second moments matrix $\widetilde C(4)$ at location~$i$ applying four measurements is calculated as
(in units of mm and mrad)
\begin{equation}
\label{overlineC_i}
\widetilde C(4)=
\begin{bmatrix}
8.6 &  -4.3 & -3.2 & -1.1 \\
-4.3&  3.4 &  -0.8 & 1.5 \\
-3.2& -0.8  & 11.2 & -3.1\\
-1.1 &  1.5&  -3.1 & 1.9
\end{bmatrix}.
\end{equation}

\begin{figure}[hbt]
\centering
\includegraphics*[width=85mm,clip=]{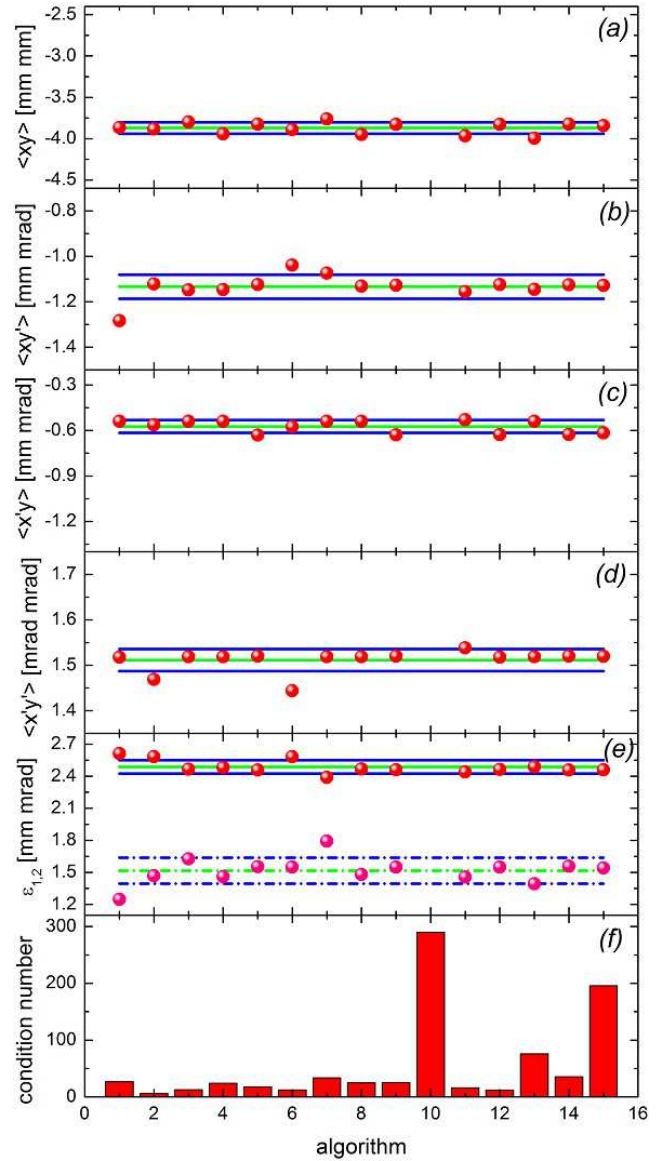}
\caption{Reconstructed results at location~$i$ applying six measurements with the skew triplet being switched on. The green line indicates the mean value and the blue lines indicate the $\pm\sigma$ error range.
(a) to (d): Coupled second moments: $\langle xy \rangle$=-3.9$\pm$0.1~mm~mm, $\langle xy' \rangle$=-1.1$\pm$0.1~mm~mrad, $\langle x'y \rangle$=-0.6$\pm$0.0~mm~mrad, and $\langle x'y' \rangle$=1.5$\pm$0.0~mrad~mrad.
(e): Eigen-emittances: $\varepsilon_1$=2.5$\pm$0.1~mm~mrad, and $\varepsilon_2$=1.5$\pm$0.1~mm~mrad. (f): Condition numbers for each algorithm. The results of algorithm~$\#{10}$ are considered as unreliable for the large condition number.}
\label{large_coupling_6}
\end{figure}

The 15 evaluations of correlated second moments at location~$i$, their corresponding
eigen-emittances, and their condition number for each algorithm using six measurements
(settings $a$ and $b$ at 0$^{\circ}$/90$^{\circ}$/30$^{\circ}$) are shown in Fig.~\ref{large_coupling_6}.
The corresponding beam second moments matrix $\widetilde C(6)$ at location~$i$ applying six measurements is calculated as
(in units of mm and mrad)
\begin{equation}
\label{overlineC_i}
\widetilde C(6)=
\begin{bmatrix}
8.6 &  -4.3 & -3.9 & -1.1 \\
-4.3&  3.4 &  -0.6 & 1.5 \\
-3.9& -0.6  & 11.2 & -3.1\\
-1.1&  1.5&  -3.1 & 1.9
\end{bmatrix}.
\end{equation}

\begin{figure}[hbt]
\centering
\includegraphics*[width=85mm,clip=]{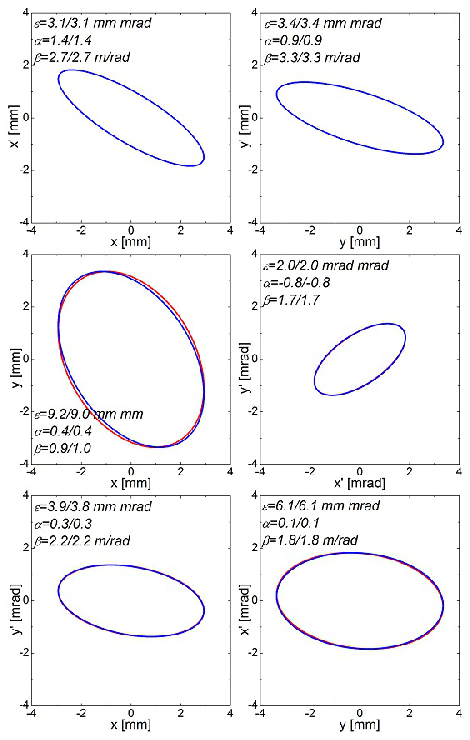}
\caption{Projected rms-ellipses from measurements applying four/six measurements with the skew triplet being switched on. The red and blue ellipses indicate the beam matrices $\widetilde C(4)$ and
$\widetilde C(6)$. The projected rms-emittances and the Twiss parameters are indicated.}
\label{coupling_large_c}
\end{figure}

Evaluation of the two eigen-emittances of $\widetilde C^(4)$/$\widetilde C(6)$ reveals $\varepsilon_1$=2.4/2.5~mm~mrad
and $\varepsilon_2$=2.1/1.5~mm~mrad. The corresponding coupling parameters $t$ are 1.2/1.8.
The beam is significantly coupled. Comparing the beam matrices $\widetilde C(4)$ and $\widetilde C(6)$, the difference between their larger eigen-emittances
is small but the smaller eigen-emittances are different.
The projected rms-ellipses of these matrices are shown in Fig.~\ref{coupling_large_c}. According to the rms-ellipses in the projections,
the rms-ellipses look very similar but feature different eigen-emittances and coupling parameters.

The larger uncertainty on the measured eigen-emittances, especially $\varepsilon_2$, for a beam inhabiting considerable correlations is already known from conventional emittance measurements in one single-plane (see appendix). The uncertainty of the measured single plane emittance is larger if the beam is strongly convergent or divergent, i.e. if it is correlated. This is just from the fact that the final observable, i.e. the emittance, is calculated from a difference (see Equ.~\ref{eqdd}) between measured quantities. Differences are much more prone to errors from their constituents as sums or products. This sensitivity known from single-plane emittance measurements occurs in four-dimensional measurements as well, as the eigen-emittances, especially $\varepsilon_2$, are also calculated from differences of measured quantities (Equ.~\ref{eqhh}). Therefore a way to reduce the error on the measured emittance is to reduce the beam correlations prior to the emittance measurements. This method has been applied successfully for single-plane measurements. In order to apply it to four-dimensional emittance measurements it needs to be demonstrated that the measured data are sufficiently accurate to perform this reduction of correlations.
\section{decoupling prospect}
\begin{figure}[hbt]
\centering
\includegraphics*[width=85mm,clip=]{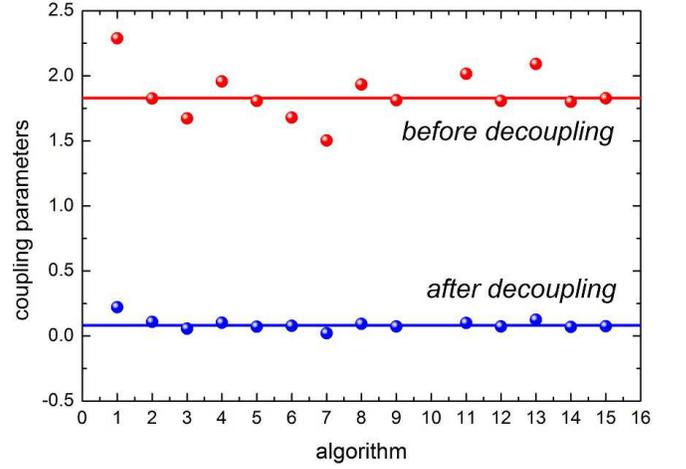}
\caption{Red and blue dots indicate the coupling parameters of matrices of $\widetilde C(6)_j$ and $\widetilde C(6)^d_j$, and solid lines indicate the
eigen-emittances of averaged beam matrices $\widetilde C(6)$ and $\widetilde C(6)^d$.}
\label{eigen_emittances_11}
\end{figure}

Any arbitrary beam line including at least three inter-plane coupling elements may serve to remove all inter-plane correlations. Here a beam line composed of a
skew quadruplet enclosed by two normal quadruplets is chosen. If this beam line is set to decouple the beam matrix $\widetilde C(4)$ calculated from four
measurements of the large coupling case, the corresponding decoupling transport matrix $R(4)$ is determined from the required gradients as (in units of mm and mrad)
\begin{equation}
\label{M_d}
R(4)=
\begin{bmatrix}
-1.2408 &-1.6164 & -0.4174 & -0.0863\\
-0.0762& -1.0492 &  0.0298 & 0.4344 \\
-0.1489 &  0.4850  &-0.5914&-2.1406 \\
0.2555 & 0.3684 & 0.5530&  0.0085
\end{bmatrix},
\end{equation}
and the decoupled second moments matrix gets
\begin{equation}
\begin{aligned}
\label{M_d}
\widetilde C(4)^d=R(4)  \widetilde C(4) {R(4)}^T
=\begin{bmatrix}
2.1 &0.0 & 0.0 & 0.0 \\
0.0&  2.1 &  0.0 & 0.0 \\
0.0 & 0.0  &2.4 &0.0  \\
0.0 & 0.0 & 0.0&  2.4
\end{bmatrix}.
\end{aligned}
\end{equation}
In the following this transfer matrix $R(4)$ is applied to each of the beam matrices $\widetilde C(6)_j$ calculated from the 15 algorithms
\begin{equation}
\begin{aligned}
\label{Cf1}
\widetilde C(6)^d_j=R(4) \widetilde C(6)_j {R(4)}^T, ~j=1,2,\cdots 15
\end{aligned}
\end{equation}
in order to test its decoupling capability. The coupling parameters before and after decoupling for each algorithm are plotted in Fig.~\ref{eigen_emittances_11}.

The coupling parameters of matrices $\widetilde C(6)^d_j$ are lower than 0.1 for each algorithm, i.e. the beam is practically decoupled.
The decoupling transfer matrix $R(4)$, constructed from four measurements will decouple all cases from six measurements with reasonable condition numbers.
Accordingly, even for beams being considerably coupled, just four measurements are required to determine the four-dimensional beam parameters with sufficient precision to allow for elimination of all inter-plane correlations by an appropriate beam line.
\section{Conclusion}
A new method using an rotatable slit/grid emittance measurement device called ROSE has been developed and commissioned to measure the four-dimensional second order beam matrix. It will allow precise and mobile four-dimensional emittance measurements without additional elements. This unique set-up works with high reliability.
During ROSE commissioning, it was found that three of the parameters extracted from the four-dimensional
beam matrix (eigen-emittance $\varepsilon_{1,2}$ and the $t$-parameter) are sensitive even to small errors in the measurements.
Despite careful choice of the optics reducing this sensitivity, fluctuations in $\varepsilon_{1,2}$ and $t$ were observed for a beam with considerable correlation. This observation confirms results from earlier single-plane emittance measurements, that featured larger errors in case the beam was correlated.
However, ROSE can provide as major deliverable the optics to fully decouple a correlated beam. This optics is quite insensitive to the exact value of $\varepsilon_{1,2}$ and $t$ as it just depends on the second moments. The latter were measured with sufficient precision. ROSE therefore can provide the input for advanced coupled beam dynamics methods as the four-dimensional beam envelope modell~\cite{Qin1,Chung1,Qin2,Qin3,Chung2,Chung3}.
\begin{acknowledgments}
One of the authors, Chen Xiao, would like to express his sincere thanks to Peter Forck at GSI for fruitful discussions.
\end{acknowledgments}
\appendix
\section{rms emittance error}
Equ.~\ref{eqdd} states
\begin{equation}
\varepsilon=\sqrt {\langle xx \rangle \langle x'x' \rangle - \langle x x' \rangle^2}=\sqrt{\frac{\langle xx \rangle \langle x'x' \rangle}{1+\alpha^2}}
\end{equation}
and because $\langle xx \rangle$, $\langle xx' \rangle$, and $\langle x'x' \rangle$ are independent, the total error of measured emittance $\delta \varepsilon$ is written to be
\begin{equation}
\begin{aligned}
&\delta \varepsilon=\\
&\sqrt{ \left(\frac{\partial \varepsilon}{\partial \langle xx \rangle} \delta \langle xx \rangle\right)^2+\left(\frac{\partial \varepsilon}{\partial \langle x'x' \rangle} \delta \langle x'x' \rangle\right)^2+\left(\frac{\partial \varepsilon}{\partial \langle xx' \rangle} \delta \langle xx' \rangle\right)^2}
\end{aligned}
\end{equation}
where $\delta \langle xx \rangle$, $\delta \langle xx' \rangle$, and $\delta \langle x'x' \rangle$ are the errors of the measured second moments. Accordingly
\begin{equation}
\begin{aligned}
&\frac{\delta \varepsilon}{\varepsilon}=\\
&\sqrt{\frac{1+\alpha^2}{ \langle xx \rangle \langle x'x' \rangle} }
\frac{\sqrt{\left( \frac{1+\alpha^2}{\beta} \varepsilon\delta \langle xx \rangle\right)^2+( \beta \varepsilon \delta \langle x'x'
\rangle)^2+4( \alpha \varepsilon \delta \langle xx' \rangle)^2}}{2 \varepsilon}
\\
&=\frac{1}{2\varepsilon}\sqrt{\left( \frac{1+\alpha^2}{\beta} \delta \langle xx \rangle\right)^2+( \beta \delta \langle x'x'
\rangle)^2+4( \alpha \delta \langle xx' \rangle)^2}.
\end{aligned}
\end{equation}

Large $\alpha$ will cause large $\frac{\delta \varepsilon }{\varepsilon}$ during the emittance measurement. In turn $\frac{\delta \varepsilon }{\varepsilon}$ is minimized for $\alpha$=0, i.e. for an uncoupled beam.
\bibliography{C.Xiao}
\end{document}